\documentclass{article}
\usepackage{deluxetable}
\usepackage{amssymb}

\usepackage{times,epsfig}
\usepackage{graphics,graphicx}
\usepackage{amsmath}
\usepackage{amsfonts}
\usepackage{psfig}

\usepackage{mathrsfs}
\usepackage{amssymb}
\usepackage{bm}
\newcommand{\affil}[1]{$^{\rm #1}$}
\usepackage[authoryear]{natbib}
\bibpunct{ (}{)}{;}{a}{}{,}
\usepackage{graphicx}
\date{} 
\newcommand{\kms}{\mbox{km\,s$^{-1}$}}
\def\kms {\ifmmode{{\rm ~km~s}^{-1}}\else{~km~s$^{-1}$}\fi}

\def\lsun {\ifmmode{{\rm ~L}_\odot}\else{~L$_\odot$}\fi}

\def\deg {^{\circ} }

\def\mjybm {\,mJy/beam}
\def\ujy {\,$\mu$Jy}

\newbox\grsign \setbox\grsign=\hbox{$>$} \newdimen\grdimen \grdimen=\ht\grsign
\newbox\simlessbox \newbox\simgreatbox
\setbox\simgreatbox=\hbox{\raise.5ex\hbox{$>$}\llap
 {\lower.5ex\hbox{$\sim$}}}\ht1=\grdimen\dp1=0pt
\setbox\simlessbox=\hbox{\raise.5ex\hbox{$<$}\llap
 {\lower.5ex\hbox{$\sim$}}}\ht2=\grdimen\dp2=0pt

\def\lsim{\mathrel{\rlap{\lower4pt\hbox{\hskip1pt$\sim$}}
    \raise1pt\hbox{$<$}}}                
\def\gsim{\mathrel{\rlap{\lower4pt\hbox{\hskip1pt$\sim$}}
    \raise1pt\hbox{$>$}}}                

\def\apj {{\it Ap.~J.}}
\def\apjl {{\it Ap.~J.\ (Letters)}}
\def\apjs {{\it Ap.~J.\ Suppl.}}
\def\aj {{\it A.~J.}}

\def\mnras {{\it MNRAS}}

\def\nat {{\it Nature}}
\def\pasa {{\it PASA}}

\def\aj{AJ}                   
\def\apj{ApJ}                 
\def\apjl{ApJ}                
\def\apjs{ApJS}               
\def\mnras{MNRAS}             

\def\nat{Nature}              
\def\iaucirc{IAU~Circ.}       

\def\grtsim{\mathrel{\hbox{\rlap{\hbox{\lower2pt\hbox{$\sim$}}}\raise2pt\hbox{$>$}}}} 
\def\lesssim{\mathrel{\hbox{\rlap{\hbox{\lower2pt\hbox{$\sim$}}}\raise2pt\hbox{$<$}}}}

\def\lsim{\,\lower2truept\hbox{${<\atop\hbox{\raise4truept\hbox{$\sim$}}}$}\,}
\def\gsim{\,\lower2truept\hbox{${> \atop\hbox{\raise4truept\hbox{$\sim$}}}$}\,}
\def\simlt{\mathrel{\rlap{\lower 3pt\hbox{$\sim$}}
        \raise 2.0pt\hbox{$<$}}}
\def\simgt{\mathrel{\rlap{\lower 3pt\hbox{$\sim$}}
        \raise 2.0pt\hbox{$>$}}}

\title{\large\bf\flushleft {The Life and Times of the Parkes-Tidbinbilla Interferometer}}
\author{\parbox{\textwidth}{\flushleft
\vspace{-0.5cm}
{\it
Ray P.\ Norris\affil{1} \&
M. J. Kesteven\affil{1}\\
{\small \affil{1}\,CSIRO Astronomy \& Space Science, PO Box 76, Epping, NSW 1710, Australia}\\
{\small \affil{}\,Email: Ray.Norris@csiro.au}
 \\
\vspace{0.4cm}
}}}
\begin{document}
%
\begin{changemargin}{.8cm}{.5cm}
\begin{minipage}{.9\textwidth}
\vspace{-1cm}
\maketitle
%
%

{\bf Abstract: 
The Parkes-Tidbinbilla took advantage  of a real-time radio-link connecting the Parkes and Tidbinbilla antennas to form the world's longest real-time interferometer. Built on a minuscule budget, it was an extraordinarily successful instrument, generating some 24 journal papers including 3 Nature papers, as well as facilitating the early development of the Australia Telescope Compact Array. Here we describe its origins, construction, successes, and life cycle, and discuss the future use of single-baseline interferometers in the era of SKA and its pathfinders.
 
}


\medskip
\medskip
\end{minipage}
\end{changemargin}

]

\section{Introduction}
 The Parkes - Tidbinbilla Interferometer (PTI) used the 64-m radio telescope at Parkes, NSW together with the 70-m DSS43 antenna of the NASA Deep Space Network at the Canberra Deep Space Communication Complex (CDSCC) at Tidbinbilla. With a baseline of 275 km, it was the world's longest real-time interferometer, and operated at frequencies of 1.6, 2.3, 6.7, 8.4, and 12.2 GHz to give angular resolutions between 0.13 and 0.02 arcsec. Because of the large antennas, it also had a high sensitivity (rms $\sim$ 0.2 \mjybm\ in 5 minutes). The PTI was conceived as part of the development of the Australia Telescope Compact Array (ATCA) \citep{frater}, both as a test-bed to tackle some of the technical challenges, and also to construct a reference frame of calibrators for the ATCA. It was successful in both these roles, but also proved to be  a powerful astronomical instrument in its own right, offering a combination of high resolution, high sensitivity, and the practical advantage of real-time correlation and data processing.
 
 At the time of its conception in 1983, the technique of Very Long Baseline Interferometry (VLBI) was being used  to connect major radio telescopes around the world, including Parkes and Tidbinbilla \citep[e.g.][]{preston83}, achieving milliarcsec resolution that could not be achieved with any other technique, yielding valuable insights into the astrophysics of Galactic and extragalactic sources. However, the VLBI observations were technically challenging, recording data on to band\-width-limiting video tapes that had to be synchronised to within a  microsecond, and which did not yield any observational data until the tapes were correlated some  weeks or months after the observations. As a result, while VLBI observations had great scientific value, they were generally limited to teams of ``black-belt'' VLBI experts. The longest real-time radio interferometer in 1983 was the Manchester University's Multi-Element Radio-Linked Interferometer Network (MERLIN) with a maximum baseline of 127 km, and which, with a  wider user base than VLBI, had made significant contributions to both Galactic and extragalactic science \citep[e.g.][]{booth81, browne82}.
 
The development of the PTI started in 1983 when a radio link was installed by NASA between Parkes and Tidbinbilla to facilitate the use of the two antennas to track the Voyager spacecraft on its flyby of Uranus.  The Chief of the Division of Radiophysics, R.H. ``Bob'' Frater. asked for expressions of interest to use the link for astronomical purposes. On 19 February 1984 a ``Preliminary Proposal'' \citep{norris84a} was submitted to construct a ``Parkes-Tidbinbilla Interferometer" at minimal cost. The proposal was evaluated, at the request of Bob Frater, by Alec Little, who supported it, resulting in a full proposal submitted in June 1984 \citep{norris84b}.  John D. Murray agreed to build the one piece of necessary hardware, the coarse delay unit (at a total cost of $\sim$\$1k), and Mike Kesteven agreed to work with Norris on building the interferometer. This small team was later augmented by Kel Wellington and Mike Batty.

First fringes were obtained on the first day of observations in a limited mode (PTI Phase 1) in June 1985, and then the system was upgraded to use two 5 MHz bands (Phase 2) using the 1024-channel Parkes correlator. It was subsequently upgraded again (PTI Phase 3) in 1992 to use the new AT correlator.

\section{Design Principles}
A conventional interferometer typically brings together the signals from two antennas, translates them to a baseband frequency, and correlates them in a special-purpose correlator. In addition, a delay must be inserted into the signal path of one arm of the interferometer, to remove group delay and bring the wavefronts from the two antennas into synchronisation, and the phase of one arm must be rotated to remove phase variations. Because of the rotation of the earth, both phase and delay must be adjusted continuously, typically requiring special purpose hardware.

The goal of the PTI was to build an interferometer with as little specialised hardware as possible. It was called a ``software interferometer'' because it used software for most of the functionality that would be provided by hardware in a traditional interferometer. It did so in three ways.
\begin{itemize}
\item Correlator: The existing Parkes autocorrelation spectrometer \citep{ables75} was used as a correlator.  
By measuring a full cross-correlation function over many delay channels, and allowing the peak of the cross-correlation function to drift through the delay window of the spectrometer, most of the fine delay and phase corrections could be performed in software after correlation rather than in hardware before correlation. Only coarse phase and delay corrections, needed to prevent decorrelation during the integration time, needed to be applied before correlation.


\item Delay Compensation:
Because the cross-correlation function is allowed to drift through the delay window of the spectrometer, sufficient delay must be applied prior to correlation to keep the peak of the cross-correlation function within the delay window of the spectrometer. In PTI Phase 1 the delay could be changed very infrequently (e.g. once every 20 minutes) and so manual adjustment was sufficient. The increased bandwidth of later phases of the PTI required the delay to be changed once per integration (i.e. every few seconds), and so a computer-addressable delay was necessary. A delay unit to achieve this was the only special-purpose hardware that had to be constructed for the PTI.

\item Phase rotator:
Because the fine phase corrections were applied after correlation, the only phase rotation needed prior to correlation was the phase rate needed to prevent decorrelation during the integration. Over a period of one second, this phase rate is approximately constant,  corresponding to a fixed frequency.
The online software calculated the phase difference between the two signals from the two arms of the interferometer. Because this phase is changing at a typical rate of tens of Hz, the phase rotation was achieved by incorporating  a  programmable phase-continuous synthesiser into the local oscillator chain, and changing its frequency every second.

\end{itemize}

A further requirement is that the oscillators at the two antennas must be coherent. All local oscillators at Parkes were locked to a Rubidium frequency standard, while those at  Tidbinbilla were locked to a Hydrogen Maser. The  Parkes Rubidium oscillator limited the coherence time (i.e. the maximum  interval between phase calibrators) to typically 10 minutes at 2.3 GHz. 

In this and in several other respects, the techniques used by the PTI resembled those used in VLBI rather than those used in other real-time interferometers.
 
\section{History}

 \subsection{PTI Phase 1}

\begin{figure}[h]
\begin{center}
\includegraphics[width=7cm, angle=0]{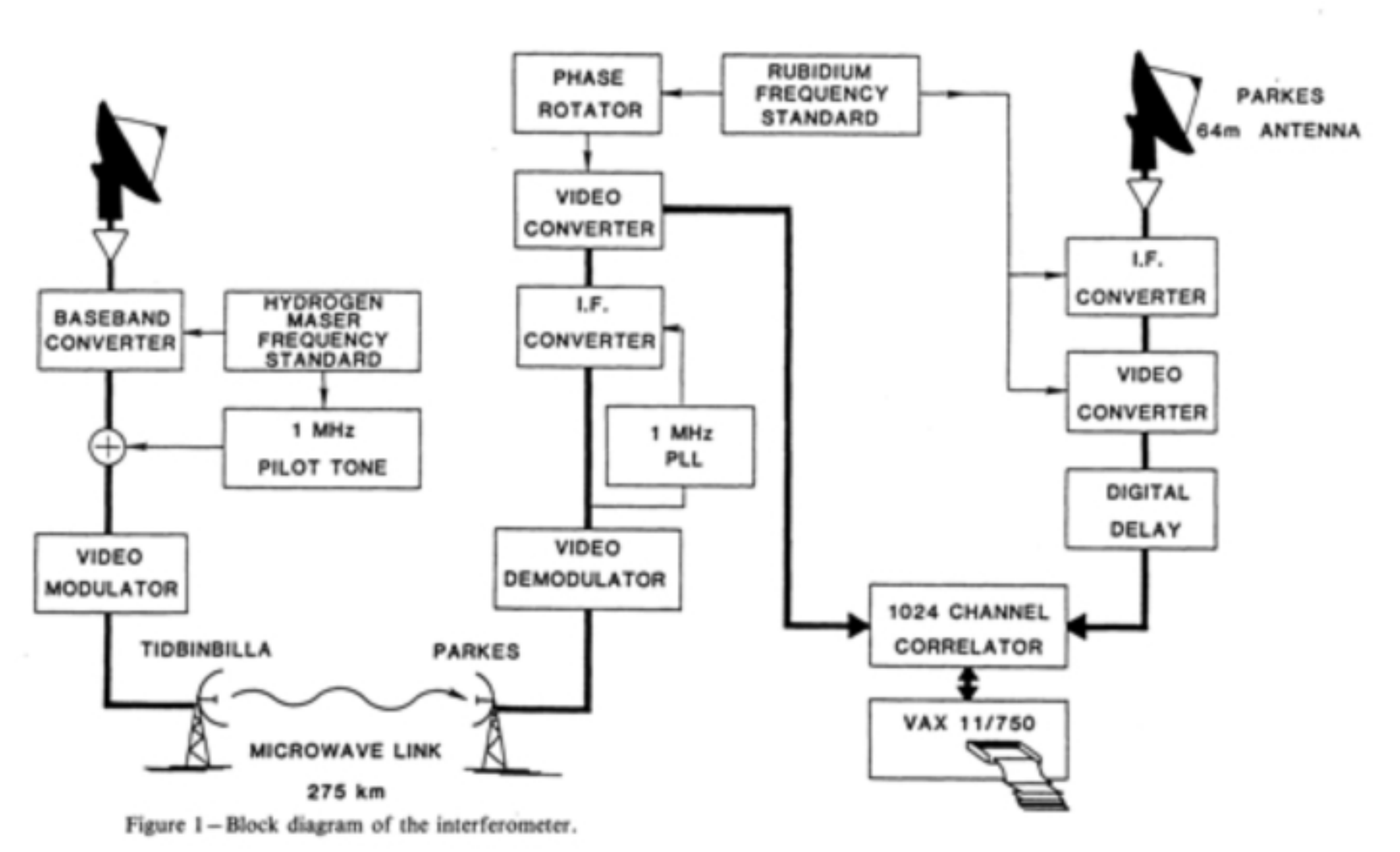}
\caption{Block diagram of PTI Phase 1, taken from \cite{norris85}.  
}
\label{pti1}
\end{center}
\end{figure}

\begin{figure}[h]
\begin{center}
\includegraphics[width=7cm, angle=0]{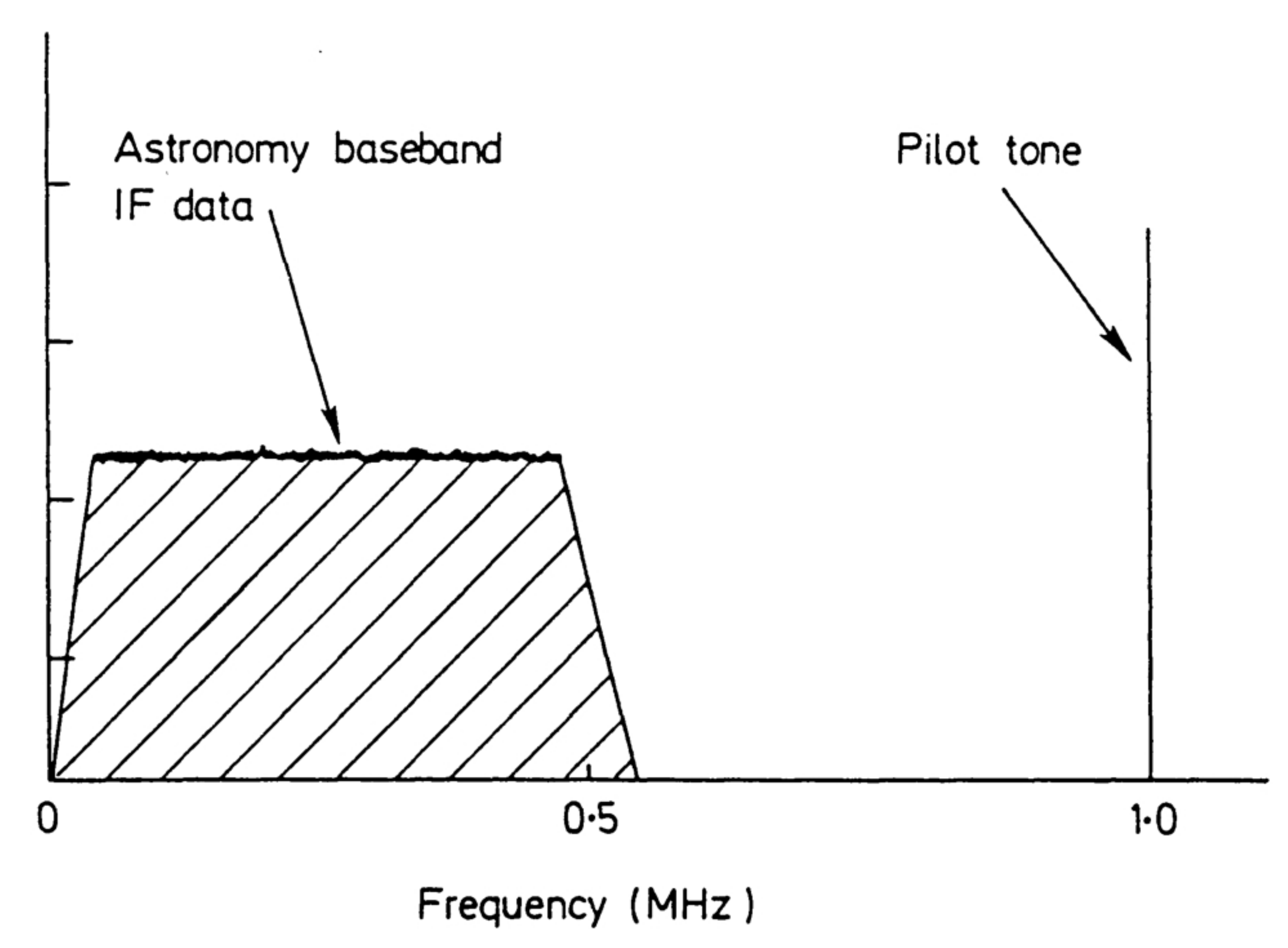}
\caption{Spectrum of the signal sent from Tidbinbilla to Parkes in PTI Phase 1. The radio-astronomy band occupies 0.5 MHz bandwidth, and a reference signal is inserted at 1 MHz to compensate for phase variations caused by instabilities in the radio link.
}
\label{tone}
\end{center}
\end{figure}

PTI Phase 1 was a proof-of-concept instrument built at very low cost. Although its bandwidth was only 0.5 MHz, the size of the antennas at each end (64m and 70m respectively) meant that its sensitivity was sufficient to do cutting-edge science. It had four novel features as follows:
\begin{itemize}
\item To remove arbitrary delay and phase variations which might occur in the radio link from  Tidbinbilla to Parkes, a 1 MHz pilot signal was inserted into the astronomy signal at Tidbinbilla, and subsequently subtracted from the astronomical signal to cancel out phase variations caused by the radio link (see Fig. \ref{tone}). Unknown to us, the radio link already had a phase compensation system, and so this part of the system turned out to be redundant, and was removed for PTI Phase 2. 

\item To remove the phase changes caused by Earth rotation PTI used an off-the-shelf frequency synthesiser (a Rockland Model 5100, operating at about 1.25 MHz), whose frequency was updated every second. To prevent stochastic long-term drifts, the phase of this synthesiser was also set to zero at the end of each integration cycle time of 10s, causing phase steps which were then removed in software. Because frequency (and thus phase) changes could only be made at intervals of one second, the data also contained a residual phase error which varied slowly enough that it could be removed in software.

\item To remove the changes in group delay caused by Earth rotation, together with the 1.3ms delay  of the radio link from Tidbinbilla to Parkes,  we performed a delay correction in three stages. 
\begin{enumerate}
\item A coarse delay unit with a minimum delay step size of 256 $\mu$s used a simple shift register to remove the bulk of this delay, and was typically adjusted by hand at the start of each scan (typically 20 minutes), 
\item The online software calculated the position of the peak of the cross-correlation function in the correlator, and  transferred only 512 channels centred on that peak position, effectively shifting the cross-correlation function to remove the residual delay by an integral number of delay channels.
\item The residual fine delay variations then appeared as a phase gradient across the frequency spectrum, which again could be removed in the online software by applying a phase gradient to the complex spectrum.
\end{enumerate}

\item 
Because telescope time on the two largest antennas in the Southern hemisphere was at a premium, we constructed a ``dummy telescope" to simulate every aspect of the PTI. A simulated noise spectrum could be injected into the spectrometer, and all the online and offline processing performed on this simulated data. The use of this simulator considerably speeded the debugging and commissioning of the interferometer.
\end{itemize}

\begin{figure}[h]
\begin{center}
\includegraphics[width=7cm, angle=0]{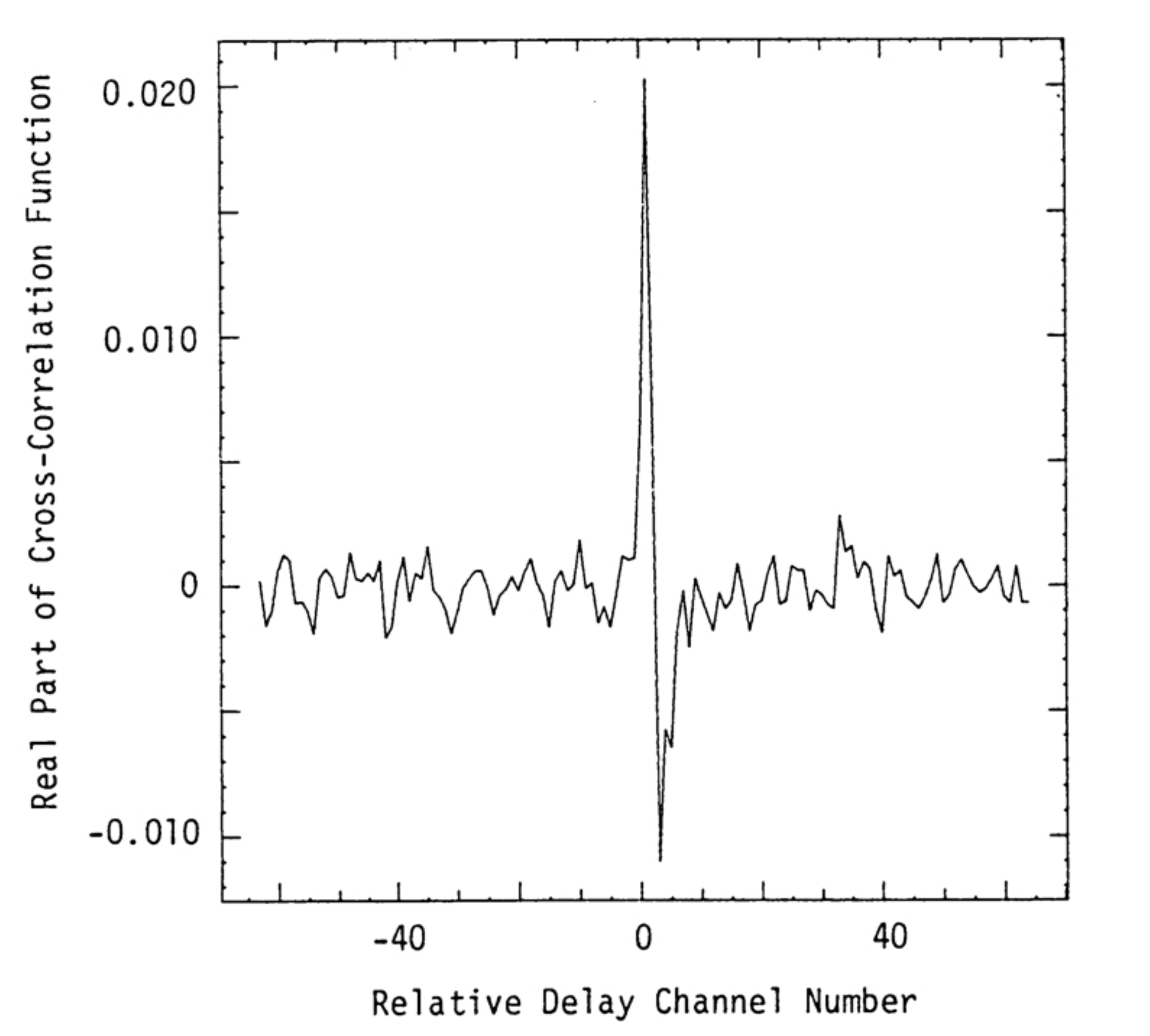}
\caption{``First fringes'' recorded with the PTI on 27 June 1985, on the quasar 3C273, taken from \cite{norris85}. The plot shows the real part of the correlated power as a function of delay. 
}
\label{fringe}
\end{center}
\end{figure}

After exhaustive testing and debugging using the dummy telescope,  the PTI hardware and software were first connected to the real antennas on 27 June 1985,  and fringes on the source 3C273 (shown in Fig. \ref{fringe}) were obtained  on that same day. Subsequent observations with PTI Phase 1 focussed on improving phase stability by understanding the phase variations introduced by the hardware, starting on the planned program of calibrator observations, and preparing for the higher bandwidth PTI Phase 2.

 \subsection{PTI Phase 2}
In 1986, PTI  was upgraded to use two channels of 5MHz each, providing a factor of 10 increase in bandwidth over PTI Phase 1. The resulting PTI Phase 2 \citep{norris88a}  was in active operation from 1986 to 1993. The greater bandwidth required the correlator to use a faster sampling clock (10 MHz), which in turn meant that the delay space covered by the 500 channels of the correlator was only 50 $\mu$s. To keep the peak of the cross-correlation function within the correlator window, a new computer-addressable digital delay unit was built, enabling the delay to be adjusted every integration period (typically 5 seconds). However, the Rockland synthesiser continued to provide phase rotation.

\begin{figure}[h]
\begin{center}
\includegraphics[width=7cm, angle=0]{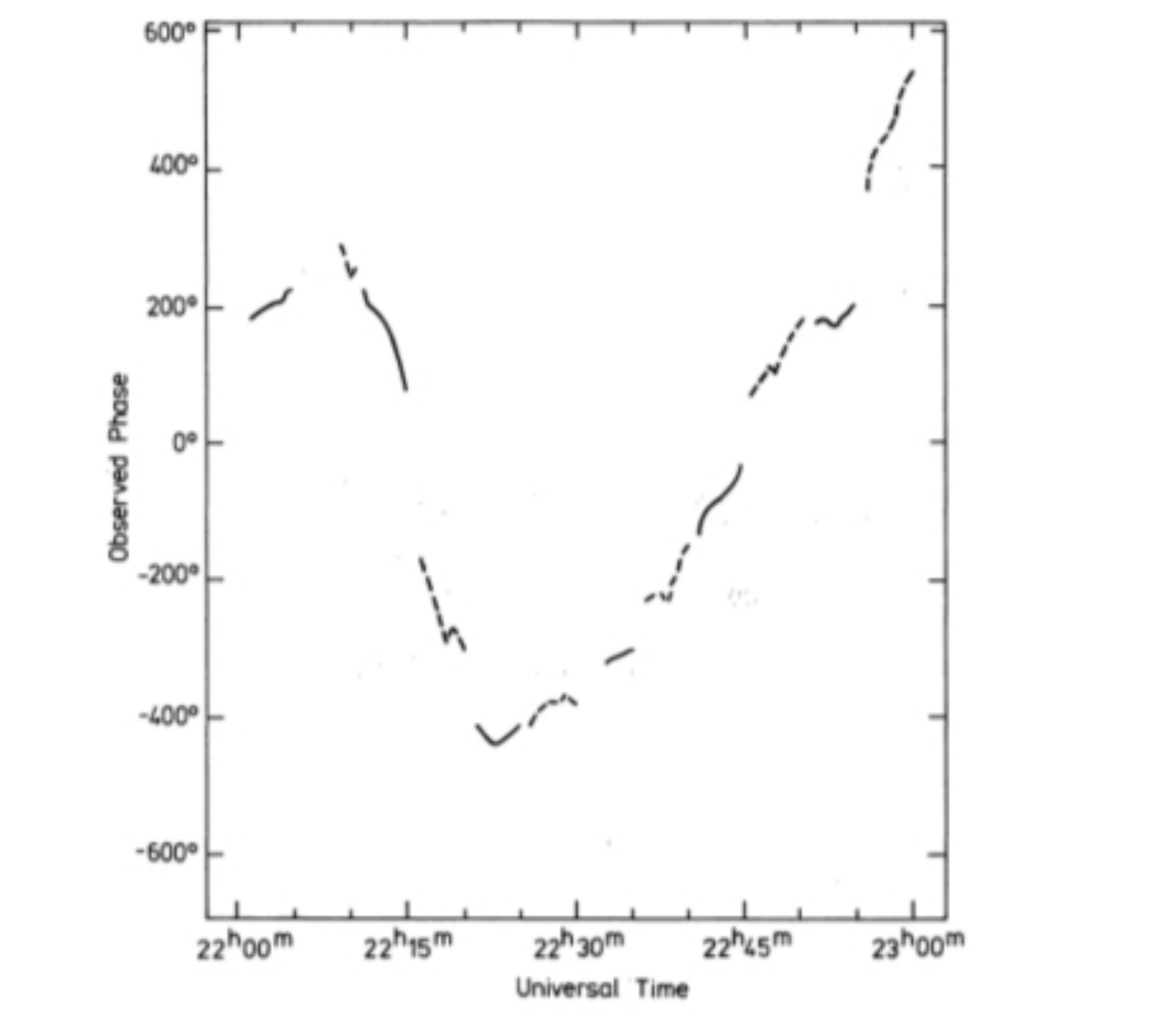}
\caption{A plot of interferometer phase at 2.3 GHz as a function of time for two unresolved calibrators (0414-189 and 0434-188) separated by 5 degrees in the sky. The observed phase variations are caused by a combination of ionosphere, troposphere, and instability of the Rubidium frequency standard used at Parkes.
}
\label{phase}
\end{center}
\end{figure}

At this time, arguably the most productive period of the PTI, the operational parameters were those shown in Table 1. At 1.7 GHz, the PTI achieved a sensitivity of 1 mJy in 15 minutes with a resolution of 0.13 arcsec. Longer effective integration times were achieved by switching against a phase-reference calibrator source, which could be as far as $5 \deg$ away with no significant loss of phase (see Fig \ref{phase}).

\begin{table}[h]
\begin{center}
\caption{PTI Phase 2 Specifications}
\label{specs}
\begin{tabular}{ll}
\hline
Baseline & 275 km  \\
Available bandwidths & 0.5, 1.0, 2.0, 5.0, 10.0 MHz \\
Frequency resolution & 256 complex spectral channels \\
Operating frequency & 1.7, 2.3, 8.4 GHz \\
Resolution (fringe spacing) & 0.13, 0.09, 0.03 arcsec\\
Typical coherence time & 15,10,3 minutes\\
Rms sensitivity in a coherence time & 1, 1.5, 3 mJy \\
\hline
\end{tabular}
\medskip\\
\end{center}
\end{table}

PTI Phase 2 was used for a wide  range of scientific work including pulsar proper motions, mapping of OH and methanol masers, searching for radio cores in quasars, active galaxies, and infrared galaxies. 
\begin{figure}[h]
\begin{center}
\includegraphics[width=7cm, angle=0]{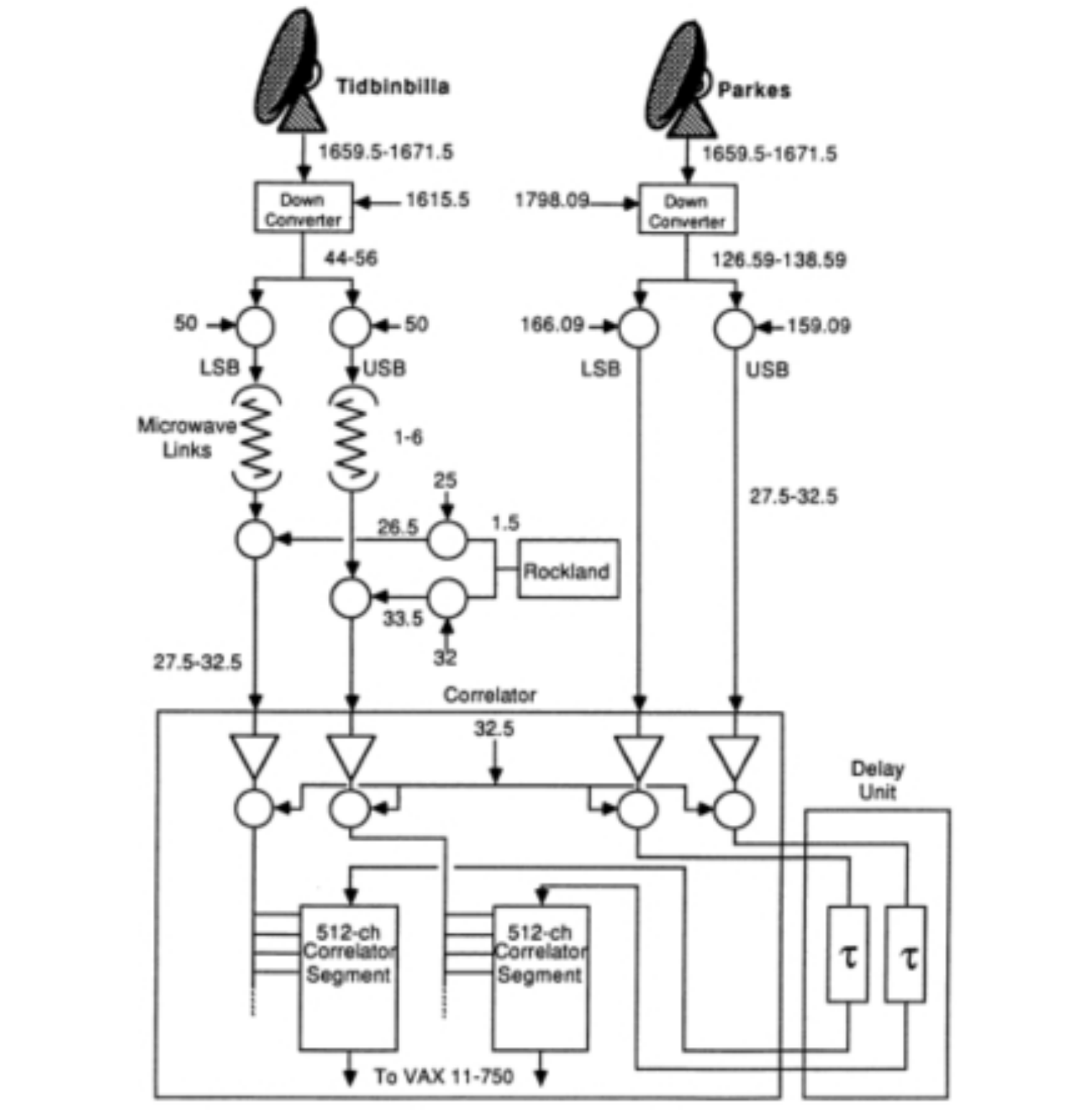}
\caption{The block diagram of PTI Phase 2.  The Parkes signal path uses the Standard Parkes frequency translators, although Marconi synthesisers rather than Systron Donners are necessary for phase stability.
}
\label{pti2}
\end{center}
\end{figure}

\subsection{PTI Phase 3}
In 1992 a new correlator was installed at Parkes, making the PTI the only observing mode to require the old correlator. Upgrading the PTI to use the new correlator would mean that (a) the old correlator could then be scrapped, saving on maintenance, and (b) the PTI could benefit from the higher reliability and better performance (wider bandwidth, 2-bit sampling) of the new correlator. 
In June 1992 a proposal was submitted to upgrade the PTI \citep{norris92} to use the new AT correlator. The upgrade was approved and in 1993/4 the PTI was significantly revised \citep{norris93} to use the new correlator, and at the same time a number of other changes were made to increase the sensitivity (by nearly a factor of 2), increase the versatility, and make it more user-friendly. The PTI continued to operate in this mode until about 1998, when the performance of VLBI recording systems overtook the capabilities of the PTI.

In PTI Phase 3, the IF signal from the Tidbinbilla antenna was converted to baseband at Tidbinbilla and limited to two bands of 0-8 MHz. At Parkes, the two baseband signals from Tidbinbilla were converted up to 8-16 MHz IF by a conversion chain which includes two programmable Stanford Research Systems frequency synthesisers operating at about 16 MHz. The frequency of these synthesiser was adjusted every second by the on-line software to provide the phase rotation. This process was restarted at the start of each scan, so that the absolute phase information was lost between scans, but was later reconstructed by the online software.  The signal from Parkes was converted to two 8-16 MHz bands using a standard down-conversion scheme. 

All four 8-16 MHz inputs from Parkes and Tidbinbilla were two-bit sampled, and then the Parkes signal was delayed by an amount equal to the geometric path difference between the two signals.  This delay was changed only once every integration, and the change in delay due to the earth's motion during the 5-second integration time led to a small amount of decorrelation. The amount of this decorrelation was corrected offline (so that amplitude calibration was corrected), but obviously the lost signal-to-noise ratio could not be recovered.

\section{Science}
\subsection{Overview} During its period of operation, some  24 journal papers, including 3 Nature papers (summarised in Table \ref{papers}) were published based on PTI data, together with many conference papers. The PTI was also used as a real-time check for VLBI observations. In this section we discuss some of the science highlights.

\subsection{SN1987A}

SN 1987A, in the Large Magellanic Cloud, was discovered on 24 February 1987 \citep{kunkel87} by Shelton and Duhalde at the Las Campanas Observatory in Chile, and by Jones in New Zealand. Easily visible to the naked eye, it was the closest observed supernova since SN 1604, and the nearest by far since the invention of radioastronomy. When news of it reached CSIRO on 25 February, Dick Manchester convened an urgent meeting of the astrophysics group to discuss what observations we should make. Since single-dish continuum observations would probably not have the sensitivity to detect it, there were three potential observing modes to make the historical first radio observations: pulsars (championed by Manchester), the Tidbinbilla short-baseline interferometer (championed by Dave Jauncey), and the recently-upgraded PTI (championed by Norris). It was decided that all three should observe it, with time on the antennas shared equally between them. We were also aware that the Sydney University group planned to observe it with the Molonglo Observatory Synthesis telescope (MOST).

Norris drove up to Parkes that evening while the Parkes staff ejected the scheduled observers and installed the 2.3 GHz receiver. Unfortunately, the receiver did not cool down until the next day, and so the first Parkes observations of SN1987A didn't take place until 26 February. The PTI observations showed a detection within a few minutes of acquiring the source, giving us not only the flux density but also an upper limit on the size. The next day the flux density was measured with the PTI at 8.4 GHz, and the light-curve was followed over the next few days until it sank below the detection limit on 5 March. The results, together with the MOST observations, were then published in Nature \citep{turtle87}, showing both the time dependence and the spectral properties.

\begin{figure}[h]
\begin{center}
\includegraphics[width=7cm, angle=0]{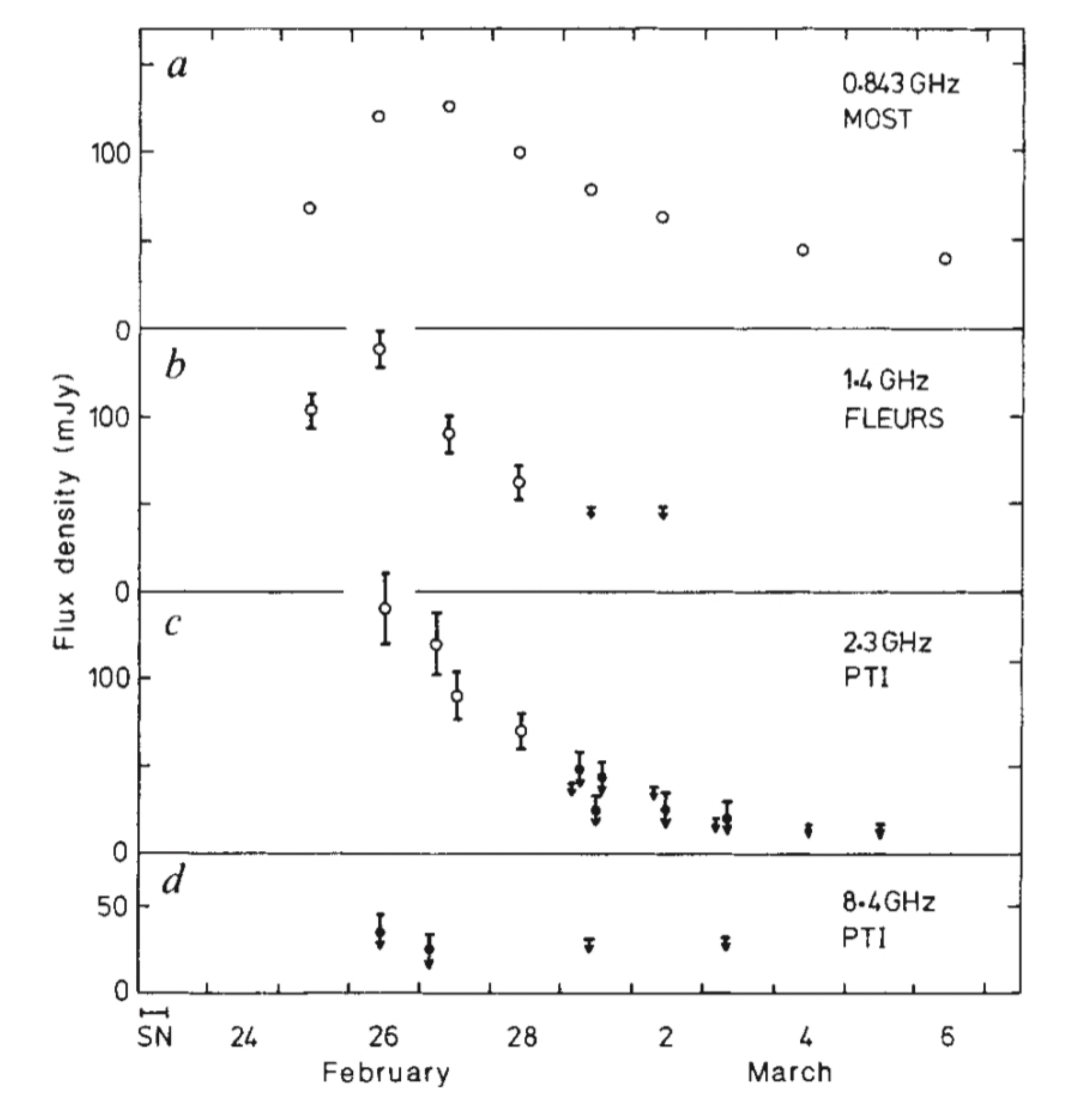}
\caption{The radio flux of supernova SN1987A plotted as a function of time at four frequencies, from \cite{turtle87}. The lower two plots show PTI data.
}
\label{sn87a}
\end{center}
\end{figure}

Scientifically, the results (Fig. \ref{sn87a}).were extremely important to our understanding of supernovae, since this was by far the  most detailed and sensitive study of the radio properties of a supernova. It was also the first major paper to result from the PTI, establishing the role of the PTI as a cutting-edge instrument able to deliver significant science results.

\subsection{Methanol Masers}
Methanol masers were first discovered by \cite{batrla87} at 12.178 GHz. OH and H$_{2}$O masers had been used for many years to study star formation regions, but the discovery of the methanol masers, which were even stronger and just as widespread as the OH masers, seemed to herald an exciting new era of maser studies of star formation. Methanol masers had not been previously discovered because they lay well away from radio-astronomy bands, at a frequency which was swamped, in the US and Europe, by interference from TV broadcasting satellites.

Norris refereed the Batrla et al. paper, and wrote a short piece \citep{norris87a} emphasising the importance of their discovery. He also obtained  the permission of Batrla et al. to propose observations of the masers with Parkes and PTI prior to their paper appearing in Nature. Kel Wellington bought cheap consumer satellite TV receivers and mounted them on the Parkes and Tidbinbilla telescopes, and within four weeks of the Nature publication  we had found the methanol masers were very common in star formation regions \citep{norris87b}. A few months later the PTI observations showed that the  masers were typically distributed over a few arcsec surrounding ultra-compact HII regions \citep{mccutcheon88, norris88b}. Those first PTI images were the source of significant anxiety because we were unsure of the sign of the phase, and an incorrect sign would cause the images to be inverted. Despite conducting many tests to check the phase polarity, we were still uncertain when the paper was submitted. Fortunately, the maps subsequently turned out to be correct. 

We noted in our first paper  that the methanol masers appeared to be in lines. This has since been confirmed by many authors, although the cause of these lines is still a matter of dispute with both edge-on disks \citep{norris98} and outflows \citep{walsh98} being potential causes.

\begin{figure}[h]
\begin{center}
\includegraphics[width=7cm, angle=0]{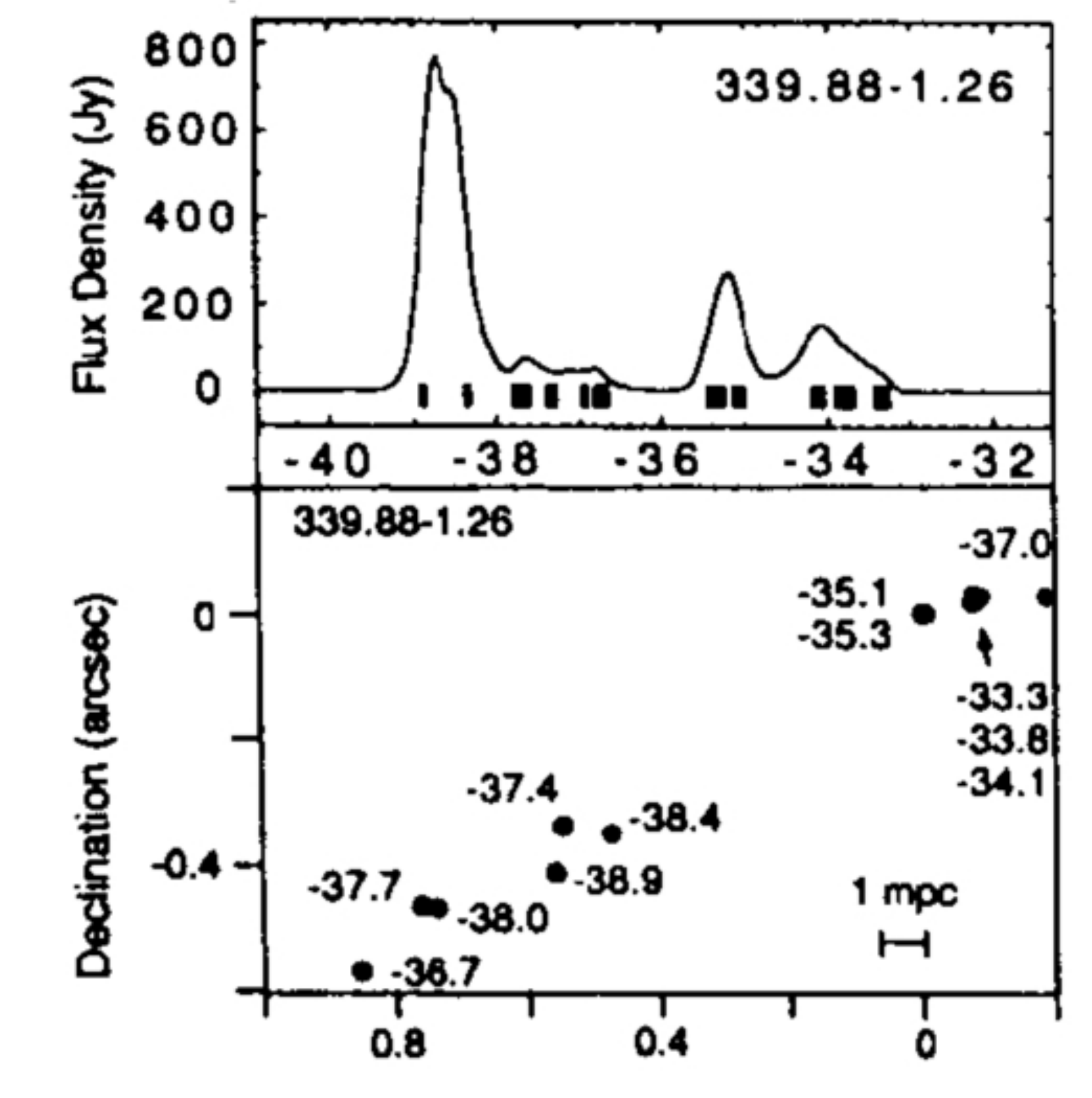}
\caption{A PTI map of the 12.2 GHz methanol masers in the source G339.88-1.26, from \cite{norris88b}. (top) the spectrum, with the horizontal axis showing the lsr velocity in km/s. (bottom) the map of the masers, shown as spots, with velocities in km/s marked against them. This map shows that the masers are arranged in a linear sequence, which subsequent observations have shown to be common in these masers, although the cause of this is still a matter of controversy.
}
\label{masers}
\end{center}
\end{figure}

\subsection{Pulsar Proper Motions}
In 1988, the proper motions of 27 pulsars had been measured, showing the pulsars to have surprisingly high velocities,  in some cases exceeding the escape velocity of the Galaxy. However, almost all these measurements had been made by one group \citep{lyne82}, using the same instrument and technique, and so it was important to make an independent check on these results. Furthermore, projecting the velocity vector backwards would enable the origin of the pulsar to be found, enabling us to check the assumption that their origin should lie at the center of a supernova remnant. 

Although PTI could not measure absolute positions of sources, it could measure the position of a source relative to a nearby calibrator with great accuracy, and so could be used to make a direct measurement of pulsar proper motions. It could also measure parallax, enabling the distance to a pulsar to be measured independently of any model-dependent assumptions, testing the models used to estimate the distance to pulsars using dispersion measure.

The first pulsar to be tackled with the PTI was the Vela Pulsar. It was one of only four pulsars known to be associated with a supernova remnant (SNR), offering a rare chance to understand the relationship between pulsars and their progenitors. \cite{bailes89} used the PTI to measure its proper motion in a two-stage process. First, the PTI was used to observe potential phase reference calibrators near the Vela pulsar. The nearest suitable source was 40 arcmin away, so that the PTI had to be used in a switching mode with a 4-minute cycle, with 2 minutes on the pulsars followed by 2 minutes on the reference. Since NASA security precluded the Parkes observers from driving the Tidbinbilla antenna directly, this required the Tidbinbilla operators to manually switch the antenna every two minutes. Nevertheless, the technique was successful, and Bailes et al. measured the proper motion of the pulsar to an accuracy of 5 milliarcsec/year. Surprisingly, they showed that, although the pulsar was associated with the Vela SNR,  it did not originate in the centre of the SNR, implying that the SNR had expanded asymmetrically since its birth.

The PTI was then used to measure the proper motions of six other pulsars. Again, a two stage process was used, and in this case pulsars were chosen which had phase reference calibrators within the primary beam, removing the need for switching. The results  \citep{bailes90a} essentially confirmed the Lyne et al. result that pulsars are high-velocity objects migrating from the Galactic Plane. In one case, the proper motion showed that a pulsar did not originate from the centre of the SNR that had been assumed to be its progenitor, but may have come from a nearby young stellar cluster.

Finally, the PTI was used to measure the parallax of a pulsar to an accuracy of 0.3 milliarcsec \citep{bailes90b}, twice as accurate as any previous measurement, enabling an independent calibration of the pulsar distance scale which was based on dispersion measure.

\begin{figure}[h]
\begin{center}
\includegraphics[width=7cm, angle=0]{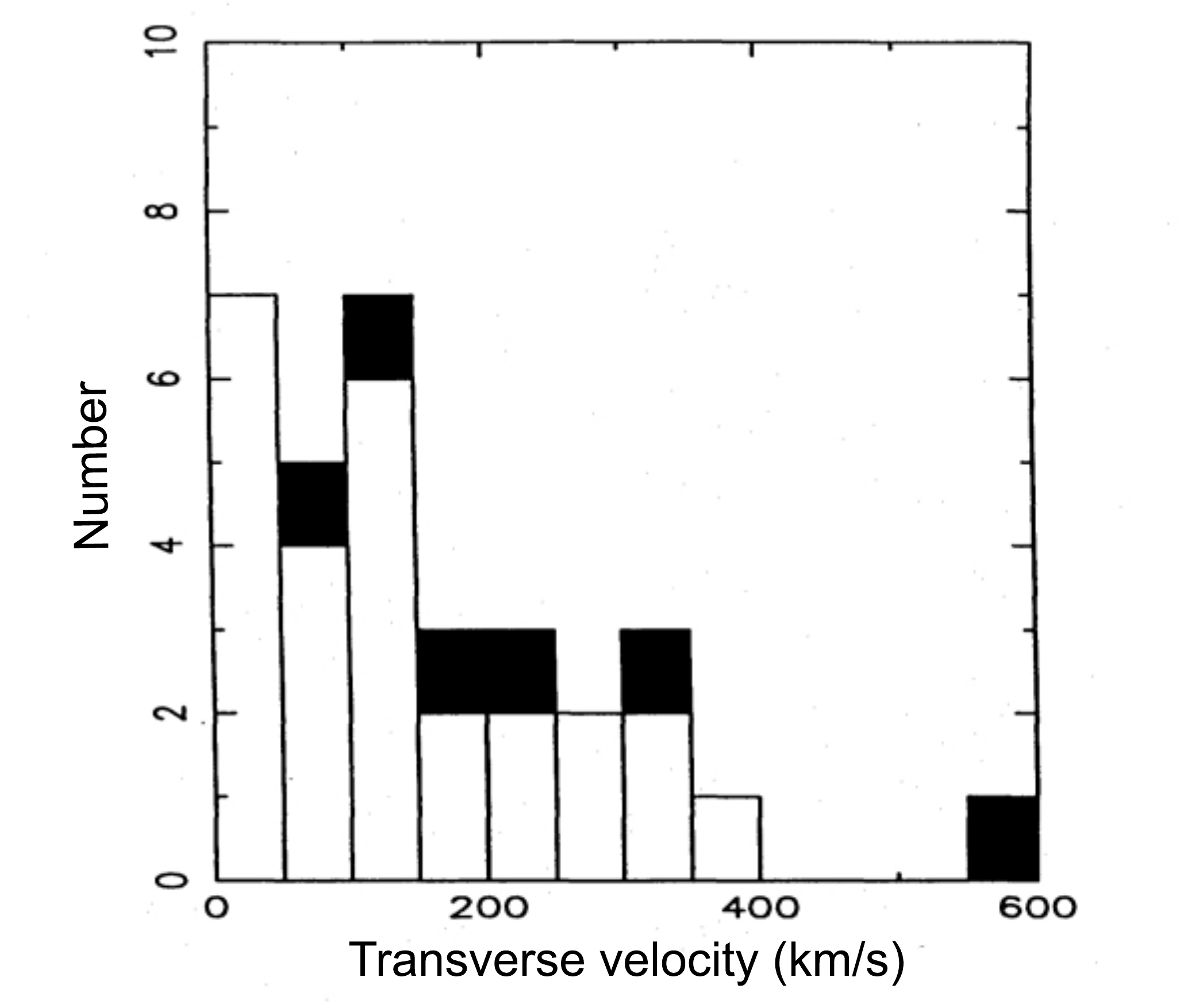}
\caption{The distribution of transverse velocity of pulsars derived from PTI and other interferometric observations, from \cite{bailes90a}.
}
\label{pulsars}
\end{center}
\end{figure}

\subsection{Calibrators}
A significant part of the initial justification for building the PTI was to construct a reference frame of calibrators fro the ATCA. This program was the first to be started, by Norris and Kesteven,  when the PTI commenced operations in 1985, and continued throughout the first few years of PTI. As both Norris and Kesteven became heavily committed to the development of the ATCA, this project was taken over by Bob Duncan, who brought it to a successful conclusion eight years later \citep{duncan93}.

\begin{figure}[h]
\begin{center}
\includegraphics[width=7cm, angle=0]{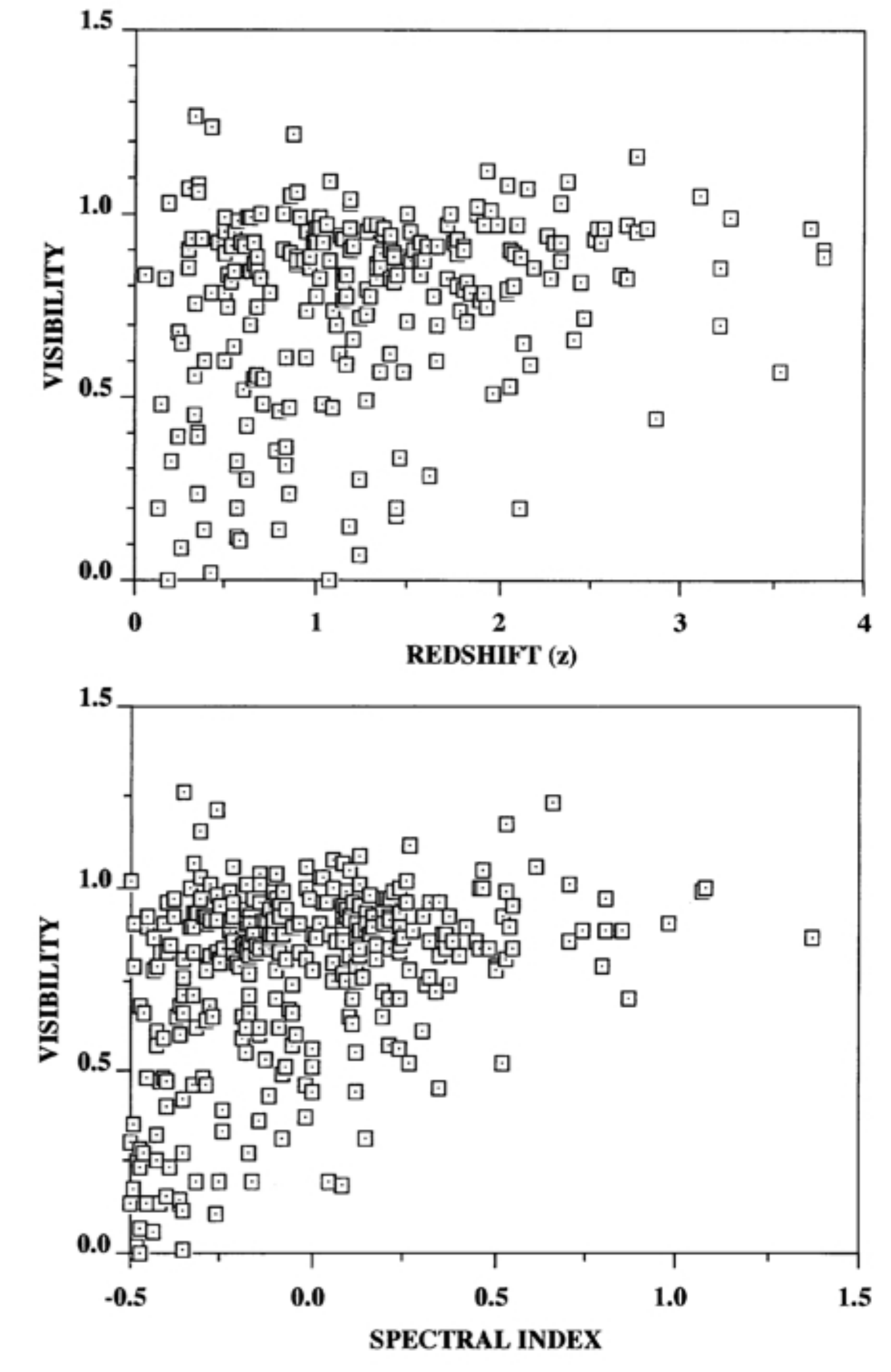}
\caption{(top) Visibility (defined as PTI flux/single-dish flux) of the putative calibrator sources as a function of redshift, showing that high-redshift objects tend to be more compact, from \cite{duncan93}. Above z=1 the angular size of a standard ruler is approximately independent of distance, so this plot appears to show an evolutionary effect.(bottom) Visibility  as a function of spectral index, showing that compactness increases with spectral index - the least compact sources tend to have steeper spectra.}
\label{calibrators}
\end{center}
\end{figure}

As well as identifying potential calibrators for the ATCA, this project had an unexpected astrophysical outcome.
In 1985, several studies had indicated that flat-spectrum sources tended to be compact, but no definitive study of a large sample had been conducted at milliarcsec resolution. To construct a calibrator catalogue, the PTI was used to survey all sources from the PKS survey \citep{bolton79} with a spectrum flatter than -0.5, to identify those that were unresolved and therefore potentially good calibrators. The program was successful in identifying calibrators, but perhaps more importantly showed a clear correlation between spectral index and compactness, and with redshift (see Fig. \ref{calibrators}). This result was unexpected because the redshift range is sufficiently high that cosmic curvature causes the angular size of a standard ruler to be roughly independent of redshift. The PTI result therefore implied that there is a real evolutionary effect: in a flux-limited sample, radio-loud sources at high redshift are physically smaller than  low redshift sources. The reason for this is yet to be satisfactorily explained, but is presumably a result of cosmic evolution of radio-loud AGN.

\subsection{Compact cores in AGNs}
Since AGNs have a much higher brightness temperature than star-forming (SF) galaxies (i.e. they can generate much strong radio emission from a small region of a given size), they are much more easily detectable with high-resolution instruments such as PTI. The PTI was therefore a powerful tool for distinguishing whether  radio sources were powered by AGN or SF activity. Although a non-detection would not necessarily rule out an AGN in any individual case, it should be possible to distinguish clearly between the powering mechanisms of samples of sources.

As the  PTI was first starting operation, a number of groups \citep{aaronson84, allen85, houck85} had independently discovered that a number of bright IR sources, at that time named either ELF (Extremely Luminous FIR source) or ULIRG (Ultra-Luminous IR galaxy), had unexpectedly high luminosities, rivalling those of quasars. However, it was unclear  whether they were powered by SF or AGN. These were therefore the first target for the PTI compact-core work, and \cite{norris88c, norris90} showed that while some contained obscured AGN, most did not, instead being powered primarily by SF activity.

\cite{slee90, slee94} used the same technique to examine a sample of optically-selected early-type galaxies  with radio emission, and found radio cores similar to those seen in radio galaxies. While the PTI flux was strongly correlated with the VLA core flux, it was only weakly correlated with any of the larger-scale indicators, at either  radio or optical wavelengths. Surprisingly, sources with a weaker total power appeared to be more core-dominated. A similar result was obtained by \cite{jones94} who observed a sample of extended radio galaxies chosen from the Molonglo Reference Catalogue and found only a weak correlation between the PTI core power and the total power. These latter results supported the unified models (e.g. \cite{padovani90}) in which PTI core flux would be determined primarily by orientation, while large-scale emission would be independent of this. \cite{sadler99} showed that the cores in bright ellipticals  are qualitatively similar to the central engines of radio galaxies but far less powerful, while those in spiral galaxies are qualitatively different, providing clues to why radio galaxies are always ellipticals rather than spirals.

The unified model for Seyfert galaxies was tested by PTI observations of a large sample of Seyfert galaxies by \cite{roy94}. Surprisingly, the PTI detection rate was significantly higher for Sy2 galaxies than Sy1 galaxies, contrary to the prediction of the unified models, although the result was uncertain because of the small sample size. \cite{sadler95} extended this work by using the PTI at two frequencies to observe a heterogeneous sample of spiral galaxies, finding that those  parsec-scale cores that were present tended to be steep-spectrum, in contrast to the flat-spectrum cores seen in higher-power radio galaxies. \cite{roy98} subsequently observed a large sample of Seyferts with the PTI and found, surprisingly, that most followed the FIR-radio correlation (FRC), suggesting that while the radio cores were undoubtedly AGN, much of the radio emission from the Seyfert was generated by star-forming activity. For those galaxies which do not lie on the FRC, it was found that the difference between the total radio flux and the PTI core flux followed the FRC significantly better than the total radio flux, confirming that much of the radio emission was generated by SF. This result was confirmed in a  different sample by \cite{hill01}, who found that SF dominates the luminosity of composite galaxies, but that even a minor AGN contribution can be distinguished by optical spectroscopy.

\cite{morganti97} applied the PTI to a complete flux-density-limited sample of strong radio sources, and found that the PTI core flux is strongly correlated with the  radio flux from the core region, but only weakly correlated with the total flux from the source. They also showed that the properties of the two Fanaroff-Riley types \citep{fanaroff} differed, suggesting that FRI cores are less strongly beamed than those in FRII galaxies. Their observations were broadly consistent with the unified model, but showed a number of complexities not reflected in the simplest models.

\cite{heisler98} used the PTI to observe a well-defined sample of IRAS galaxies with warm far-infrared colours and found that, contrary to expectations, none of those with SF optical spectrum contained a buried AGN, while those with AGN spectra appeared to represent nascent AGN growing in power on their way to becoming a radio galaxy.

In all the work cited above it was assumed that the detection of a core signified an AGN. This was tested by \cite{kewley99,kewley00} who combined PTI observations of an IR-selected sample of galaxies with spectroscopy and careful theoretical modelling. They showed that a clump of radio supernovae in a  SF galaxy could generate a PTI detection, mimicking an AGN. However, the incidence of such radio supernovae is small, and the luminosities smaller than most AGN, so that in most cases a PTI detection can still be taken to imply an AGN.

\section{PTI and  the Australia Telescope Compact Array}

A primary motivation for building PTI was to help build a calibrator catalogue for the ATCA, and it clearly succeeded in that role. As noted above, its value in generating new science was probably even greater. However, its contribution to the construction and commissioning of the ATCA itself is, in hindsight, also significant. 

First, it provided a hands-on focus for those who were given the job of designing the software for the ATCA. Without that, their role would have been to write memos for an interferometer whose commissioning date might be some years away. The PTI offered an opportunity to test the ideas and expertise in practice. Even now, software modules written for PTI continue to be used as part of the ATCA software. For example, the RPFITS format which is still used for ATCA data was first developed for the PTI.

Second, it provided a valuable tool for commissioning and debugging the ATCA. For example, the first image produced by the ATCA \citep{norris90b} was generated in AIPS from uv data which had been manipulated, edited, and calibrated in the PTI data reduction program, PTILOOK. 

\section{Conclusion}
It is surprising that the PTI was such a successful instrument, given its low budget, and the fact that conventional VLBI techniques could, in principle, achieve many of the results obtained with the PTI.

We suggest the following factors were responsible for its success:

\begin{itemize}
\item A perception (not necessarily accurate) that VLBI could only be used by ``black-belt 
VLBI gurus'', whilst PTI could be used by anyone. This positive perception of PTI was helped by a user manual which guided the user through all observing and data reduction steps, and a dummy telescope so that users could practice on dummy data prior to their observations.

\item Real-time results encouraged large surveys of many sources with short observations on each. In principle, VLBI could also be used in this mode, but in practice rarely was, perhaps because of a fear that a complicated schedule might cause errors that would not be detected until correlation took place, perhaps months later.

\item Real-time results encouraged rapid publication. The data were usually reduced while still observing, and the first drafts of some of the  observational PTI papers cited here were generated during the observations, while the observers were still engaged and motivated. While this can happen in other branches of astronomy, it is not, of course, possible with VLBI.

\end{itemize}

The PTI was a specialised instrument, unable to produce images, but capable of doing one job very well: measuring complex visibilities on a long baseline with high sensitivity. While of limited value for studying individual sources, such instruments are of enormous value for studying large samples of sources. 

SKA and its pathfinders such as ASKAP \citep{johnston08} will produce catalogs of millions of sources, for which VLBI imaging will be impossible given realistic telescope time allocations. A next-generation single-baseline instrument could be built using ASKAP and the Parkes antenna equipped with a phased-array feed. With a GHz bandwidth, this interferometer would be able to survey tens of millions of sources in a few months, to \ujy\ sensitivities. Such an instrument would be a powerful resource for identifying which populations of galaxies contained AGN. 
We predict that the era of single baseline interferometers is far from over.

\section*{Acknowledgments}

We are indebted to the following people without whose help and expertise the PTI would never have taken shape:
Mike Batty,
Graeme Carrad,
Dave Cooke,
Bob Frater,
Alec Little,
Dave Jauncey,
David Loone,
John Murray,
John Reynolds,
Kel Wellington,
and of course all the staff at Parkes and at CDSCC, Tidbinbilla.

\onecolumn

\begin{table}[h]
\begin{center}
\caption{PTI Journal papers (This includes only peer-reviewed journal papers presenting PTI data, and excludes a number of  conference papers)}
\label{papers}
\begin{tabular}{lll}
\hline
Subject area & No. of papers & References\\
\hline
Instrumental & 2 & \cite{norris85, norris88a}\\
SN1987A & 1 Nature & \cite{turtle87}\\
Masers & 3 incl. 1 Nature & \cite{mccutcheon88,norris88b, zijlstra01}\\
Pulsar proper motions & 3 incl 1 Nature & \cite{bailes89,bailes90a, bailes90b}\\
Calibrators & 1 & \cite{duncan93} \\
Radio cores in AGN & 14 & \cite{norris88c,norris90,slee90}\\
& &\cite{jones94,roy94,slee94} \\
& &\cite{sadler95,morganti97, heisler98}\\
& &\cite{roy98,kewley99,sadler99}\\
& &\cite{kewley00, hill01}\\
\hline
\end{tabular}
\medskip\\
\end{center}
\end{table}


\begin{thebibliography}{}
\bibitem[Aaronson \& Olszewski(1984)]{aaronson84} Aaronson, M., \& Olszewski, E.~W.\ 1984, \nat, 309, 414 
\bibitem[Ables et al.(1975)]{ables75} Ables, J.~G., Cooper, B.~F.~C., Hunt, A.~J., Moorey, G.~G., \& Brooks, J.~W.\ 1975, Review of Scientific Instruments, 46, 284 
\bibitem[Allen et al.(1985)]{allen85} Allen, D.~A., Roche, P.~F., \& Norris, R.~P.\ 1985, \mnras, 213, 67P 
\bibitem[Bailes et al.(1989)]{bailes89} Bailes, M., Manchester, R.~N., Kesteven, M.~J., Norris, R.~P., \& Reynolds, J.\ 1989, \apjl, 343, L53 
\bibitem[Bailes et al.(1990a)]{bailes90a} Bailes, M., Manchester, R.~N., Kesteven, M.~J., Norris, R.~P., \& Reynolds, J.~E.\ 1990a, \mnras, 247, 322 
\bibitem[Bailes et al.(1990b)]{bailes90b} Bailes, M., Manchester, R.~N., Kesteven, M.~J., Norris, R.~P., \& Reynolds, J.~E.\ 1990b, \nat, 343, 240 
\bibitem[Batrla et al.(1987)]{batrla87} Batrla, W., Matthews, H.~E., Menten, K.~M., \& Walmsley, C.~M.\ 1987, \nat, 326, 49 
\bibitem[Bolton et al.(1979)]{bolton79} Bolton, J.~G., Savage, A., \& Wright, A.~E.\ 1979, Australian Journal of Physics Astrophysical Supplement, 46, 1 
\bibitem[Booth et al.(1981)]{booth81} Booth, R.~S., Norris, R.~P., Porter, N.~D., \& Kus, A.~J.\ 1981, \nat, 290, 382 
\bibitem[Browne et al.(1982)]{browne82} Browne, I.~W.~A., Clark, R.~R., Moore, P.~K., et al.\ 1982, \nat, 299, 788 
\bibitem[Duncan et al.(1993)]{duncan93} Duncan, R.~A., White, G.~L., Wark, R., et al.\ 1993, \pasa, 10, 310 
\bibitem[Fanaroff \& Riley(1974)]{fanaroff} Fanaroff, B.~L., \& Riley, J.~M.\ 1974, \mnras, 167, 31P 
\bibitem[Frater et al.(1992)]{frater} Frater, R.~H., Brooks, J.~W., \& Whiteoak, J.~B.\ 1992, Journal of Electrical and Electronics Engineering Australia, 12, 103 
\bibitem[Heisler et al.(1998)]{heisler98} Heisler, C.~A., Norris, R.~P., Jauncey, D.~L., Reynolds, J.~E., \& King, E.~A.\ 1998, \mnras, 300, 1111 
\bibitem[Hill et al.(2001)]{hill01} Hill, T.~L., Heisler, C.~A., Norris, R.~P., Reynolds, J.~E., \& Hunstead, R.~W.\ 2001, \aj, 121, 128 
\bibitem[Houck et al.(1985)]{houck85} Houck, J.~R., Schneider, D.~P., Danielson, G.~E., et al.\ 1985, \apjl, 290, L5 
\bibitem[Johnston et al.(2008)]{johnston08}Johnston, S., et al. 2008, {\it Experimental Astronomy}, 22, 151
\bibitem[Jones et al.(1994)]{jones94} Jones, P.~A., McAdam, W.~B., \& Reynolds, J.~E.\ 1994, \mnras, 268, 602 
\bibitem[Kewley et al.(1999)]{kewley99} Kewley, L.~J., Heisler, C.~A., Dopita, M.~A., \& Norris, R.~P.\ 1999, Advances in Space Research, 23, 1045 
\bibitem[Kewley et al.(2000)]{kewley00} Kewley, L.~J., Heisler, C.~A., Dopita, M.~A., et al.\ 2000, \apj, 530, 704 
\bibitem[Kunkel et al.(1987)]{kunkel87} Kunkel, W., Madore, B., Shelton, I., et al.\ 1987, \iaucirc, 4316, 1 
\bibitem[Lyne et al.(1982)]{lyne82} Lyne, A.~G., Anderson, B., \& Salter, M.~J.\ 1982, \mnras, 201, 503 
\bibitem[McCutcheon et al.(1988)]{mccutcheon88} McCutcheon, W.~H., Wellington, K.~J., Norris, R.~P., et al.\ 1988, \apjl, 333, L79 
\bibitem[Morganti et al.(1997)]{morganti97} Morganti, R., Oosterloo, T.~A., Reynolds, J.~E., Tadhunter, C.~N., \& Migenes, V.\ 1997, \mnras, 284, 541 
\bibitem[Norris (1984a)]{norris84a}Norris, R.P., 1984a,"A Preliminary Proposal for a Parkes-Tidbinbilla Interferometer", AT Technical memo 
\bibitem[Norris (1984b)]{norris84b}Norris, R.P., 1984b,"A Proposal for a Parkes-Tidbinbilla Interferometer", AT Technical memo AT/20.1.1/002
\bibitem[Norris et al. (1985)]{norris85} Norris, R.~P., Batty, M.~J., Kesteven, M.~J., \& Wellington, K.~J.\ 1985, \pasa, 6, 137 
\bibitem[Norris (1987)]{norris87a} Norris, R.~P.\ 1987, \nat, 326, 12 
\bibitem[Norris et al. (1987)]{norris87b} Norris, R.~P., Caswell, J.~L., Gardner, F.~F., \& Wellington, K.~J.\ 1987, \apjl, 321, L159  
\bibitem[Norris et al.(1988a)]{norris88a} Norris, R.~P., Kesteven, M.~J., Wellington, K.~J., \& Batty, M.~J.\ 1988a, \apjs, 67, 85 
\bibitem[Norris et al.(1988b)]{norris88b} Norris, R.~P., Caswell, J.~L., Wellington, K.~J., McCutcheon, W.~H., \& Reynolds, J.~E.\ 1988b, \nat, 335, 149 
\bibitem[Norris et al.(1988c)]{norris88c} Norris, R.~P., Kesteven, M.~J., Troupe, E., \& Allen, D.~A.\ 1988, \mnras, 234, 51P 
\bibitem[Norris et al.(1990)]{norris90} Norris, R.~P., Kesteven, M.~J., Troup, E.~R., Allen, D.~A., \& Sramek, R.~A.\ 1990, \apj, 359, 291 
\bibitem[Norris et al.(1990)]{norris90b} Norris, R.~P., Kesteven, M.~J., Sramek, R.~A., et al.\ 1990, \pasa, 8, 252  
\bibitem[Norris (1992)]{norris92}Norris, R. P., 1992, ``Proposed Upgrade to the Parkes-Tidbinbilla Interferometer'', 13 June 1992
\bibitem[Norris (1993)]{norris93}Norris, R. P., 1993, AT Technical memo AT/42.1.2/003
\bibitem[Norris et al.(1998)]{norris98} Norris, R.~P., Byleveld, S.~E., Diamond, P.~J., et al.\ 1998, \apj, 508, 275 
\bibitem[Padovani \& Urry(1990)]{padovani90} Padovani, P., \& Urry, C.~M.\ 1990, \apj, 356, 75 
\bibitem[Preston et al.(1983)]{preston83} Preston, R.~A., Wehrle, A.~E., Morabito, D.~D., et al.\ 1983, \apjl, 266, L93 
\bibitem[Roy et al.(1994)]{roy94} Roy, A.~L., Norris, R.~P., Kesteven, M.~J., Troup, E.~R., \& Reynolds, J.~E.\ 1994, \apj, 432, 496 
\bibitem[Roy et al.(1998)]{roy98} Roy, A.~L., Norris, R.~P., Kesteven, M.~J., Troup, E.~R., \& Reynolds, J.~E.\ 1998, \mnras, 301, 1019 
\bibitem[Sadler et al.(1995)]{sadler95} Sadler, E.~M., Slee, O.~B., Reynolds, J.~E., \& Roy, A.~L.\ 1995, \mnras, 276, 1373 
\bibitem[Sadler(1999)]{sadler99} Sadler, E.~M.\ 1999, Advances in Space Research, 23, 823 
\bibitem[Slee et al.(1990)]{slee90} Slee, O.~B., White, G.~L., Reynolds, J.~E., \& Caganoff, S.\ 1990, Astrophysical Letters and Communications, 27, 361 
\bibitem[Slee et al.(1994)]{slee94} Slee, O.~B., Sadler, E.~M., Reynolds, J.~E., \& Ekers, R.~D.\ 1994, \mnras, 269, 928 
\bibitem[Turtle et al.(1987)]{turtle87} Turtle, A.~J., Campbell-Wilson, D., Bunton, J.~D., et al.\ 1987, \nat, 327, 38 
\bibitem[Walsh et al.(1998)]{walsh98}  Walsh, A.~J., Burton, M.~G., Hyland, A.~R., \& Robinson, G.\ 1998, \mnras, 301, 640 
\bibitem[Zijlstra et al.(2001)]{zijlstra01} Zijlstra, A.~A., Chapman, J.~M., te Lintel Hekkert, P., et al.\ 2001, \mnras, 322, 280 
%


\end{thebibliography}
\end{document}